\documentstyle[12pt]{article} 
\begin{document}
\renewcommand{\theequation}{\thesection .\arabic{equation}}
\newcommand{\beq}{\begin{equation}}
\newcommand{\eeq}{\end{equation}}
\newcommand{\beqn}{\begin{eqnarray}}
\newcommand{\eeqn}{\end{eqnarray}}
\newcommand{\slp}{\raise.15ex\hbox{$/$}\kern-.57em\hbox{$\partial
$}}
\newcommand{\slA}{\raise.15ex\hbox{$/$}\kern-.57em\hbox{$A$}}
\newcommand{\lnA}{\raise.15ex\hbox{$/$}\kern-.57em\hbox{$A_{c}^{N}$}}
\newcommand{\slB}{\raise.15ex\hbox{$/$}\kern-.57em\hbox{$B$}}
\newcommand{\bP}{\bar{\Psi}}
\newcommand{\bC}{\bar{\chi}}
 \newcommand{\hs}{\hspace*{0.6cm}}

\title{Path-integral fermion-boson decoupling at finite temperature}
\author{M.V.Man\'{\i}as$^{a,b}$,C.M.Na\'on$^{a,b}$, and M.L.Trobo$^{a,b}$}
\date{December 1996}
\maketitle

\thispagestyle{headings}
\markright{\thepage}

\begin{abstract}

\hs We show how to extend the standard functional approach to bosonisation, 
based on a decoupling change of path-integral variables, to the case in which a
finite temperature is considered. As examples, in order to both illustrate and
check the procedure, we derive the thermodynamical partition functions for the
Thirring and Schwinger models.

\end{abstract}
 
\vspace{3cm}
Pacs: \\ 
\hspace*{1,7 cm} 11.10.-z \\
\hspace*{1,7 cm} 11.15.-q

\noindent --------------------------------

\noindent $^a$ {\footnotesize Depto. de F\'\i sica.  Universidad
Nacional de La Plata.  CC 67, 1900 La Plata, Argentina.}

\noindent $^b$ {\footnotesize Consejo Nacional de Investigaciones
Cient\'\i ficas y T\'ecnicas, Argentina.}
\newpage
\pagenumbering{arabic}

\section{Introduction}
Some time ago a path-integral approach to bosonisation was developed 
\cite{Boson}, which has been shown to be very useful in
the study of a variety of (1+1) problems such as the Kondo effect \cite{FRS},
fermion fields in topological backgrounds \cite{MNT}, 1D many-body systems
\cite{NRT}, etc. This method is based on a decoupling change of path-integral
variables, whose corresponding Fujikawa \cite{Fuji} Jacobian $J_{F}$ yields a
non-trivial contribution to the kinetic piece of the bosonised Lagrangian
density. In a relevant contribution, Reuter and Dittrich \cite{RD} have 
computed $J_{F}$ at non-zero temperature, paving thus the road for the 
application of the decoupling technique in the context of finite temperature
Quantum Field Theory (QFT) \cite{BER} \cite{MAT} \cite{DJ} \cite{TU}. 
However, only very recently this path-integral approach was employed to compute 
the fermion condensate at finite density \cite{CS}, whereas a systematic and 
detailed computation of thermodynamical quantities within the functional 
bosonisation scheme is still lacking. The principal aim of this work is to help 
filling this gap. Apart from its academic interest, the formulation we present 
here is a necessary first step towards the implementation of a 
finite-temperature functional treatment of models that try to describe some 
features of the recently developed 1D semiconductors \cite{Voit}.\\
In order to illustrate the procedure we shall consider the two most popular 
(1+1) QFT's, namely, the Thirring \cite{th} and Schwinger ($QED_{2}$) \cite{S} 
models. These theories have been extensively studied at $T\neq0$, 
following operational and functional approaches different from ours. 
In particular, the thermodynamical properties of the Thirring model were 
examined by Ruiz Ruiz and Alvarez-Estrada \cite{Ruiz} and Yokota \cite{Yoko}, 
who found contradictory answers for the corresponding partition function. 
This issue has been recently reconsidered by Sachs and Wipf \cite{SW}, whose 
result agrees with that of ref.\cite{Ruiz}. In the next Section we rederived 
the partition function for the Thirring model by using the above mentioned 
path-integral route to bosonization. Our result provides an independent 
confirmation of the validity of the expression first obtained in \cite{Ruiz}.\\
For completeness, in Section 3 we show how to get the thermodynamical partition
funtion of the Schwinger model \cite{RD} \cite{Love} within our framework. 
There is no controversy in this case, and thus our result coincides with that
of ref.\cite{Ruiz}, as expected. We end this paper by briefly summarizing our
results and commenting on possible applications of this formulation.\\


\section{The Thirring model}

In this section we consider the two-dimensional Thirring model 
at finite temperature using the imaginary time formalism developed by
Bernard \cite{BER} and Matsubara \cite{MAT}. We start from the Euclidean
partition function

\beq
Z= N N_{F}(\beta) \int_{antiper} D\bP~ D\Psi \exp \{- \int_{\beta} 
d^2x~ [\bP i\slp\Psi - \frac{g^2}{2} (\bP \gamma_{\mu} \Psi)^2]\}
\label{1}
\eeq

where $\int_{\beta} d^2x~$ means $\int_0^{\beta}dx^0 \int dx^1$ and 
$\beta=\frac{1}{k_{B}T}$  with $k_{B}$ the Boltzman's constant and 
$T$ the temperature. Here $N$ is an infinite constant which does not depend on
temperature. On the other hand, $N_{F}(\beta)$ is given by (please see the 
paragraph following eq.(\ref{17}))

\beq
ln N_{F}(\beta)=2 ln \beta \sum_{n} \int \frac{dk}{2\pi}
\label{2}
\eeq

As is well-known, the functional integral in (\ref{1}) must be extended over the 
paths with antiperiodicity conditions in the Euclidean time variable $x^0$:

\beqn
\Psi (x^0+\beta, x^1) & = & - \; \Psi(x^0, x^1) \nonumber\\
\bP~ (x^0+\beta, x^1) & = & - \; \bP~(x^0, x^1) 
\label{3}
\eeqn

Exactly as one does in the zero-temperature case, we can eliminate the quartic 
fermionic interaction introducing a vector field $A_{\mu}$ through the identity

\beq
\exp \{-\frac{g^2}{2} \int_{\beta}d^2 x (\bP \gamma_{\mu} \Psi)^2 \} \propto
\int_{per}DA_{\mu} \exp \{\int_{\beta}d^2 x [\frac{A_{\mu}^2}{2}+
g \bP\slA\Psi] \}
\label{4}
\eeq

We have to impose periodicity conditions for the bosonic $A_{\mu}$ field over 
the range $[0,\beta]$. The partition function then results

\beq
Z= N N_{F}(\beta) \int_{antiper} D\bP~ D\Psi 
\int_{per}DA_{\mu} \exp \{-\int_{\beta}d^2 x [\bP i\slp\Psi + 
\frac{A_{\mu}^2}{2} + g \bP\slA\Psi] \}
\label{5}
\eeq

Of course, the above equation can be expressed in terms of a $\beta$-dependent
fermionic determinant (satisfying the corresponding antiperiodic conditions) as

\beq
Z= N N_{F}(\beta) \int_{periodic}DA_{\mu} \det(i\slp+g\slA)
\exp \{-\int_{\beta}d^2 x  \frac{A_{\mu}^2}{2} \}
\label{6}
\eeq

Now, following the path-integral bosonization scheme \cite{Boson} we write the 
vector field $A_{\mu}$ in terms of two scalar fields in the form

\beq
A_{\mu}=-\epsilon_{\mu\nu}\partial_{\nu}\phi+\partial_{\mu} \eta
\label{7}
\eeq

and make a chiral transformation in the fermionic variables

\beqn
\Psi & = & e^{\gamma_5 \phi+i\eta} \chi \nonumber\\
\bP & = & \bar\chi e^{\gamma_5 \phi-i\eta}
\label{8}
\eeqn

so that we can decouple $A_{\mu}$ from the fermion fields in the determinant. 
The Jacobian associated to this change in the fermionic 
path-integral measure has been first computed, for the $T\neq0$ case, 
by Reuter and Dittrich \cite{RD}. Their result allows one to write

\beq
\det(i\slp+g\slA) = \det(i\slp) \; e^{ -\frac{g^2}{2\pi}
\int_{\beta}d^2 x \partial_{\mu} \phi \partial_{\mu} \phi}
\label{9}
\eeq

Let us stress that in writing this equation we are restricting the present
analysis to the gauge-invariant sector of the Thirring model. Indeed, the 
evaluation of the fermionic Jacobian requires a regularization prescription, 
which, in turn, involves an arbitrary parameter to be fixed on symmetry grounds 
(In passing we note that the implementation of a regularization scheme 
is essentially temperature independent \cite{BNF}). On the other hand 
the Thirring model is not a gauge theory and consequently one
should have a partition function depending on the regularization parameter,
reflecting the well-known existence of a family of solutions for this model 
\cite{th}. Among all these solutions we shall keep only the gauge-invariant
ones for two reasons: firstly to facilitate comparison with previous results
 \cite{Ruiz} \cite{Yoko}, and secondly because we have in mind the application of our
procedure in the context of many-body systems with charge conservation 
\cite{NRT}. However, it is interesting to point out that the present 
formulation seems to be specially appropriate to examine the 
non-gauge-invariant sectors of the Thirring theory at $T\neq0$. Such a model 
could be useful to describe an open (without charge conservation), 
finite-length many-body ensemble.\\  
At this stage we must also emphasize that, in contrast to the $T=0$ case, in which
the Jacobian associated to the change of bosonic variables (\ref{7}) plays no
relevant role due to the fact that it is field-independent, in the present case 
its contribution is crucial since it depends on temperature through a bosonic
determinant. To be specific let us now write down the corresponding change in
the bosonic measure,

\beq
DA_{\mu}=\det(-\Box) D\phi D\eta
\label{10}
\eeq

where $\Box=\partial_{\mu}\partial_{\mu}$.\\
Inserting (\ref{9}) and (\ref{10}) in (\ref{6}) one readily obtains

\beqn
Z & = & N N_{F}(\beta) Z_{0} \det(-\Box)\int_{per}D\phi
 \exp \{- \frac{1}{2}(1+\frac{g^2}{\pi}) \int_{\beta} d^2 x (\partial_{\mu} 
  \phi)^2 \} \nonumber \\ 
  & & \int_{per}  D\eta \exp \{- \frac{1}{2} \int_{\beta}d^2 x 
 (\partial_{\mu}\eta)^2 \}
\label{11}
\eeqn

where 
\beq
Z_{0}=\int_{antiper} D\bP D\Psi e^{-\int_{\beta} d^2 x \bP i\slp \Psi}
=det(i\slp) \nonumber\\
\eeq
From now on we shall disregard in (\ref{11}) the prefactors that are 
independent of both temperature and coupling constant.
Expressing the bosonic path-integrals in terms of determinants and using the 
property $ ln\; det A= tr\;ln A$, with $A$ the corresponding operator, we 
have

\beqn
lnZ & = & ln\;N_{F}(\beta) + tr\;ln(i\slp) + tr\;ln(-\Box) - \frac{1}{2}tr\;ln(1+
\frac{g^2}{\pi}) \nonumber \\
    & - & \frac{1}{2}tr\;ln(-\Box) - \frac{1}{2}tr\;ln(-\Box) 
\label{12}
\eeqn

Note that the only contribution which survives from the bosonic part of
the complete partition function is $\beta$-independent, because the pieces
containing $ln(-\Box)$, which depend on temperature, cancel each other.\\
In order to evaluate $lnZ_{0}$ we follow the pioneering work of
Bernard \cite{BER} and expand the fermionic fields $\Psi(x^0,x^1)$, which
are antiperiodic in the interval $0\leq x \leq \beta$, in a Fourier series:

\beq
\Psi(x^0,x^1)=\frac{1}{\beta}\sum\int\frac{dk}{2\pi}e^{ikx^1}e^{-i\omega_{n}
x^0} \Psi_{n}(k)
\label{13}
\eeq

where

\beq
\Psi_{n}(k)=\int dx^1 \int_{0}^{\beta} dx^0 e^{-ikx^1}e^{-i\omega_{n}x^0}
\Psi(x^0,x^1)
\label{14}
\eeq

and

\beq
\omega_{n}=\frac{(2n+1)\pi}{\beta}
\label{15}
\eeq

Taking into account that in eq.(\ref{12}) trace means 
$\sum_{n}\int\frac{dk}{2\pi}$, it is straightforward to obtain
 
\beq
lnZ=lnN_{F}(\beta)+\frac{1}{2}\sum_{n}\int \frac{dk}{2\pi} (k^2+\omega_{n}^2)
-\frac{1}{2}\sum_{n}\int\frac{dk}{2\pi} ln(1+\frac{g^2}{\pi})
\label{16}
\eeq

At this point some algebraic manipulations similar to those performed in 
ref.\cite{DJ} allows to write 

\beqn
lnZ & = & lnN_{F}(\beta)+2\int_{0}^{\infty} \frac{dk}{2\pi}[\frac{\beta k}{2}
+ln(1+e^{-\beta k})] \nonumber \\
    & - & 2 ln\beta \sum_{n}\int\frac{dk}{2\pi}-
    \frac{1}{2}\sum_{n}\int\frac{dk}{2\pi} ln(1+\frac{g^2}{\pi})
\label{17}
\eeqn

We see that if we identify $lnN_{F}=2 ln\beta \sum_{n}\int\frac{dk}{2\pi}$,
as it stays in (\ref{2}), we get

\beq
Z=Z_{FD} \exp \{-\frac{1}{2}ln(1+\frac{g^2}{\pi})\sum_{n}\int\frac{dk}{2\pi}\}
\label{18}
\eeq

where $Z_{FD}$ is the Fermi-Dirac distribution for massless electrons:

\beq
lnZ_{FD}=2\int_{0}^{\infty} \frac{dk}{2\pi}[\frac{\beta k}{2} 
+ln(1+e^{-\beta k})]
\label{19}
\eeq

Therefore the partition function for the Thirring model at finite-temperature
differs from the one corresponding to free massless fermions in an infinite 
constant which depends on the coupling parameter $g$ but not on temperature. 
This result exactly coincides with the expression first obtained in 
ref.\cite{Ruiz}, in contrast to the identification between Thirring and free
fermions partition functions claimed in ref.\cite{Yoko}. As stressed in \cite{SW},
this is a relevant difference, since it affects the value of the zero-point
pressure of the system.

\newpage

\section{ The Schwinger model}
 
Let us now study the thermodynamical partition function of the Schwinger 
model. As it is a gauge theory its quantization requires a gauge fixing
\cite{BER}, which gives rise to the appearence of the Faddev-Popov determinant.
We choose  to work in the Lorentz gauge $ \partial_{\mu} A_{\mu} =0$. We start 
from the expression\\

\beqn 
 Z_{S}& = & N N(\beta) \int_{antiper} D\bP \; D\Psi  \int_{per} DA_{\mu} \Delta_{FP}
 \delta( \partial_{\mu} A_{\mu} ) \nonumber \\ 
 & & \exp \{-\int_{\beta} d^2 x [\bP \; (i\slp + e \slA) \; \Psi
 - \frac{1}{4} F_{\mu \nu}^{2}]\}
\label{20}
\eeqn
where $F_{\mu \nu} = \partial_{\mu} A_{\nu} - \partial_{\nu} A_{\mu}$ and 
$N(\beta)$ is the same constant defined in (\ref{2}). Taking into account that
in the Lorentz gauge one has $\Delta_{FP} = det (- \Box)$, the above equation 
can be rewritten as

\beqn
Z_{S} & = & N N(\beta) det(-\Box) \int_{antiper} D\bP \; D\Psi  \int_{per} 
 DA_{\mu} \delta \; (\partial_{\mu} A_{\mu}) \nonumber\\
 &\exp & \{-\int_{\beta} d^2 x [\bP \; (i\slp + e \slA) \; \Psi
 - \frac{1}{4} F_{\mu \nu}^{2} -\frac{1}{2} (\partial_{\mu} A_{\mu})^2] \}
 \label{21}
 \eeqn

 In order to decouple the fermionic from the bosonic fields we perform
 the transformation (\ref{8}) and write the gauge field as in (\ref{7}).
 It is evident that the gauge condition $\partial_{\mu} A_{\mu} = 0$ becomes 
$\Box \eta = 0$. Thus, using the identity 
$\delta(\Box \eta) = \frac{1}{det (\Box)} \delta(\eta)$, and
taking into account the Jacobian generated by these changes of variables 
 we arrive at:\\
 \beqn
 Z_{S} & = & N \; N(\beta) \; det(i\slp ) \; det(-\Box) \int D \eta D\phi \nonumber\\
 & \exp & \{-\frac{1}{2} \int_{\beta} d^2x \phi [ -\frac{e^2}{\pi} \Box + \Box \Box]
 \phi \}
 \label{22}
 \eeqn
 In this expression we see that the $\eta$ field is completely decoupled from 
the $\phi$ field and has no dynamics. This means that its corresponding 
functional integral is an infinite $\beta$ independent constant that we will 
absorb in the normalization.\\
Taking logarithm in (\ref{22}), and keeping the terms that depend on the 
temperature and the coupling constant, we get 
 \beqn
 ln Z_{S} & = & ln \; N(\beta) + ln \; det(i\slp) + tr \;ln(- \Box) -\nonumber\\
 & - & \frac{1}{2} ln[(\frac{e^2}{\pi} - \Box) (- \Box)] 
 \label{23}
 \eeqn
 From now on we shall work in momentum space, where we have \\
 \beqn
 tr \; ln( - \Box) & = & \sum_{n}  \;\int  \; \frac{dk}{2 \pi}
 ln (\omega_{n}^{2} + k^{2}) \nonumber \\
 tr  \; ln ( \frac{e^{2}}{\pi} - \Box ) & = & \sum_{n}  \;\int  \; \frac{dk}{2 \pi}
 ln (\omega_{n}^{2} + k'^{2}) \nonumber \\ \nonumber \\
 \eeqn
 
with $k'^{2} = \frac{e^{2}}{\pi} + k^{2}$ and 
$\omega_{n}=\frac{2n\pi}{\beta}$.\\

 Hence we get
 \beqn
 ln Z_{S} & = & ln N(\beta) + \frac{1}{2} \sum_{n} \int \frac{dk}{2 \pi}
 (k^{2} + \omega_{n}^{2}) + \nonumber \\ 
 & + & \frac{1}{2} \{ \int \frac{dk}{2 \pi}
 2 ln \; senh \frac{k \beta}{2} - 2 ln \beta \sum_{n} \int \frac{dk}{2 \pi} 
 \} \nonumber \\ 
 & - & \frac{1}{2} \{ \int \frac{dk}{2 \pi}
 2 ln \; senh \frac{k' \beta}{2} - 2 ln \beta \sum_{n} \int \frac{dk}{2 \pi} 
 \}
 \label{25}
 \eeqn
 where one sees that the infinite $\beta$-dependent terms coming 
 from the evaluation of $tr \; ln (- \Box)$ and $tr \;ln (-\frac{e^{2}}{\pi} \Box
 + \Box \Box)$ cancel each other. After some more algebra one finally obtains 
 \beqn
 ln Z_{S} & = & 2 \int \frac{dk}{2 \pi} \{ \frac{k \beta}{2} + 
 ln ( 1 + e^{-k \beta}) \} + \nonumber \\ 
 & + & \int \frac{dk}{2 \pi} \{ \frac{\beta}{2}
 (k - k') + ln ( \frac{ 1 - e^{-k \beta}}{1 - e^{-k' \beta}}) \} 
 \label{26}
 \eeqn
which is in agreement with previous results \cite{Ruiz}. Notice that this 
partition function is not equal to that corresponding to the massive boson times
the one of free massless fermions, as one could have naively expected. Indeed,
there is also a factor associated to the zero-mass gauge excitation which
appears in the Lowenstein-Swieca solution for the massless Schwinger model 
\cite{S}. Due to this contribution, in the case $e=0$, one recovers the 
partition function for free fermions.
\newpage

\section{Summary and looking ahead}
In this work we have shown how to implement, at finite temperature, a by now
standard approach to the study of (1+1) QFT's, originally developed in the 
context of systems at $T=0$ \cite{Boson}. As examples, in Sections 2 and 3, we
have evaluated the partition functions for the Thirring and Schwinger models,
respectively. In both cases we were able to reobtain the results previously 
presented in the literature \cite{Ruiz}. Concerning our expression for the
Thirring thermodynamical partition function, it is of interest in itself since
there was some controversy between the first computations of Ruiz Ruiz and
Alvarez-Estrada \cite{Ruiz} and Yokota \cite{Yoko}. Our result provides an
independent confirmation, through a different method, of the answer given in
\cite{Ruiz} and \cite{SW}.\\
Besides its simplicity, one specific advantage of our technique is the 
possibility of examining the general (non gauge-invariant) sectors of the
Thirring model at $T\neq0$ in a natural way. A physical realization of this
model could be found if one examines the thermodynamical behaviour of an open,
non charge-preserving system of many particles in a finite volume. Within this 
formulation the lack of gauge invariance manifests itself through the appearence 
of an additional dimensionless parameter associated to the
regularization of the fermionic determinant which is at the root of the 
approach.\\ 
This work could be followed in several directions, ranging from the 
investigation of fermionic models in topological backgrounds \cite{MNT} to the
analysis of the Kondo effect at low but finite temperatures, according to the
lines of ref.\cite{FRS}. However, we think that the application of the present
formalism will be particularly fruitful when considering the non-local Thirring 
model \cite{NRT}, which has been recently proposed to describe 1D many-body 
systems. Work on this last subject is currently in progress and will be 
reported elsewhere.

\end{document}